# Predicting Changes in Affective States using Neural Networks

**Stina Lyck Carstensen, Jens Madsen and Jan Larsen**
Technical University of Denmark
Department of Applied Mathematics and Computer Science
Richard Petersens Plads, Building 324
2800 Kongens Lyngby, Denmark
{stlc,jenma,janla}@dtu.dk

## Abstract

Knowledge of patients affective state could prove to be crucial for health-care professionals in both diagnosis and treatment, however, this requires patients to report how they feel. In practice the sampling rate of affective states needs to be kept low, in order to ensure that the patients can rest. Furthermore using traditional methods of measuring affective states, is not always possible, e.g. patients can be incapable of verbal communications. In this study we explore the prediction of peoples self-reported affective state by measuring multiple physiological signals. We use different Neural networks (NN) setups and compare with different multiple linear regression (MLR) setups for prediction of changes in affective states. The results showed that NN and MLR predicted the change in affective states with accuracies of 91.88% and 89.10%, respectively.

## 1 Introduction

Physiological monitoring systems at hospitals are more and more often used as tools for diagnosing and treatment of patients. Signals recorded using such systems contain information about the patient' physiological state, but might also contain information about the patient' psychological well-being. An affective state is a psychological construct defining emotions and mood [1], and is hence one way of representing a psychological states [1]. Russell's circumplex model can be used to describe affective states [2]. This model spans affect in a two dimensional space, by using the valence and activation axes [3]. If affective states can be predicted from physiological signals, it would be possible to make continuously measurements of patients' affective states. Such prediction would be of great help for health-care professionals when treating as well as diagnosing patients.

Measurements of affective states are currently performed, by asking people how they feel, using a great variety of self-report methods [1]. Self-report methods are very dependent on the subjects understanding of the method, as well as the scale and vocabulary used. Furthermore when a question is repeatedly asked, to obtain e.g. affective states, people tend to drift, which in turn induces a bias [4, 5, 6]. The bias can be caused by *anchoring and adjustment*, that is people ignore their absolute position in the affective space, and instead *adjust* their answer relative to previous answer (the anchor) [4, 5, 6].

Activation of the sympathetic nervous system influences many physiological measures, such as Heart Rate (HR), Heart Rate Variability (HRV) and Electro Dermal Activity (EDA). These measures are associated with affective states[7], and are often available in clinical settings, making them ideal for predictive modeling.

In previous studies where affective states have been predicted from physiological signals, a multitude of different features have been tested [8, 9]. In these studies features such as HR, Electro Dermal Re-sponse (EDR), and Skin Conductance Level (SCL) were extracted to predict the perceived activation using linear models. However these models failed at predicting the dimension of valence. In [9] they were able to predict valence with up to $89.75\%$ accuracy (subject dependent) using non-linear neural networks(NN). Other studies have used other modalities such as facial expressions, speech features [10], self-reports of activity[11] etc. However most of these methods were not designed to be used in clinical settings, mainly due to practical issues, e.g. facial expressions are difficult to acquire and patients could be incapable of verbal communication. Furthermore, these approaches concentrate on modeling the individual making them subject dependent models and therefore rarely applicable in clinical settings [9].

### 1.1 Goal of current study

The goal of this study is to take the first step towards making subject independent predictions of *changes* in affective states that can be used in clinical settings. Previous studies have found that the valence dimension is best modeled by non-linear models[9, 12]. Comparisons of non-linear models have shown that especially non-linear NNs are very effective for predicting subject-dependent levels of valence, as well as activation, from physiological signals[13, 14]. We therefore use non-linear NNs to predict *changes* in subjects' affective state, using subject-independent models, and compare these predictions with predictions from linear subject-independent models. We use physiological signals, which are often available in clinical settings, and hence no extra equipment for monitoring e.g. facial expression or speech signals is needed. Based on the anchoring theory, and in contrast to previous studies [8, 9, 10, 11, 15], we model the changes in affective states instead of predicting each annotation independently. We thus relieve the anchoring bias making modeling more robust. Finally, we will investigate how performance is influenced by the model parameters and features used.

## 2 Methods

### 2.1 Experimental procedure

We design a laboratory experiment with three different experimental conditions. The conditions are designed to ensure changes in the experienced affective states. They included: a baseline-, a stressor-, and a music stimulus condition. The purpose of the baseline condition is to make subjects relax, the stressor condition is designed to increase the perceived level of activation, by giving the subjects a stress-full task. Since music induces a great variety of emotional responses [16] it is an appropriate stimuli to make subjects cover the affective space. 32 subjects participated in the experiment (17 women, 15 men). The subjects are not informed about the purpose of the experiment, a priory.

During each condition the EDA and the blood-volume by means of photoplethysmography (PPG) are recorded. The features extracted from the physiological signals include: SCL and EDR from EDA as well as HR, and HRV from PPG. To make physiological features comparable across subjects, each subjects feature-set is standardized.

After each of the conditions the subjects are asked to place a marker in the affective space, which represents their affective states. However, after the last two conditions, the marker they placed in the previous condition is also shown, and they are now asked to place their marker relative to how they felt before. In this way we are able to control what anchor subjects' use for their affective states, and use this knowledge to reduce the anchoring bias.

### 2.2 Models

To predict changes in perceived valence and activation, from the physiological features, we use NN. The Levenberg-Marquardt method is used for parameter estimation [17]. To explore the most optimal solution, different NN-model setups are tested including: 1) All possible feature combinations; 2) Varying the number of hidden layers and hidden units from 1 to 4 and 1 to 12, respectively, and 3) varying the amount of temporal data. We test the models using the features extracted in a window 20 seconds before the answer is given and up to 280 seconds before the answer is given, using a step-size of 20 seconds. The hyperbolic tan-sigmoid function is used as transfer-function in hidden layers[18], the output layer uses a linear function. In addition, Multiple Linear Regressions (MLR) was performed, to see if simple linear relation existed between the physiological signals and the change in valence and activation.

## 2.3 Performance evaluation

A hold-one-subject-out cross-validation scheme is used, resulting in subject independent models. Each subject reported their affective state three times, that is after the baseline, stressor and music condition, resulting in two test-samples for each model. For the NN, all training and testing setups are repeated 10 times to ensure that the final results would be representative for the model setup. The predictions of the NN and MLR are compared against 3 different benchmark predictions: 1) predicting the average of the valence and activation annotations in the training set, 2) random values drawn from a uniform distribution ranging over the possible target-values and 3) random values drawn from a normal distribution with mean 0 and a standard deviation of 6. To compare the performance of the benchmarks and the models, the average error for the predictions for all subjects was calculated for the NN and the MLR. The accuracy was calculated as the ratio between the error and the maximum possible error.

## 3 Results

In Figure 1a the answers given after the three conditions are shown. The figure shows that a large area of the affective space is used. Figure 1b shows the change between conditions, these are the target values used for predictions.

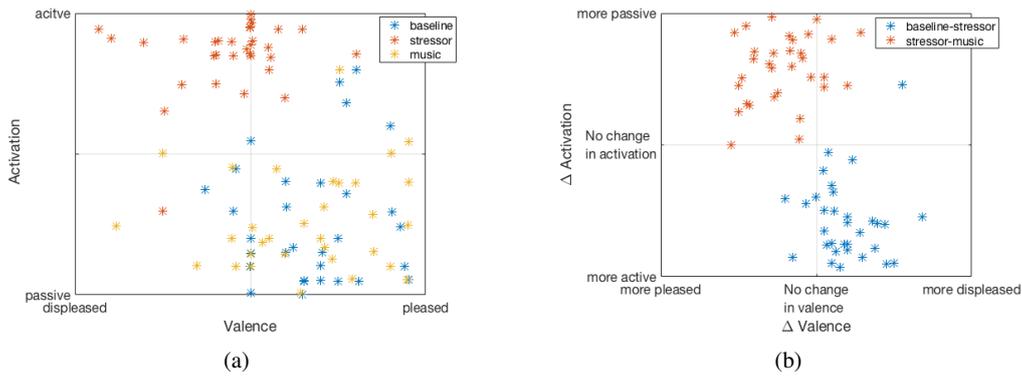

Figure 1: In Figure (a) the answers given after baseline (blue), stressor (red) and music(yellow) are shown. Figure (b) shows changes from baseline to stressor (blue) and from stressor to music(red) these are the target values for the neural network and multiple linear regressions. The experiment was performed in Danish, and the labels on the axis are hence translated from Danish to English.

The best average accuracy of predicting the valence and activation was the NN, achieving an accuracy of $91.88\%$. This network used the features EDR, HR, and HRV extracted in 260 seconds of physiological data, using 3 layers each with 11 hidden units. The best performing MLR model used the EDR and SCL extracted in 260 second windows, which resulted in an accuracy of $89.10\%$. The accuracy of best performing NN and MLR was compared using a one-way ANOVA, furthermore the Tukey-Kramer method was used as correction for multiple comparison, and the significance level was set to $\alpha = 0.05$. The comparison of the predictions from the two models, shows that the NN is not significantly better than the MLR model for predicting activation ($p = 0.68$) and valence ($p = 0.14$).

In Figure 2A the accuracy of the best performing NN and MLR models as a function of the temporal data used for feature extraction is shown. The figure also shows the accuracy of the benchmark predictions. The accuracy of the benchmark prediction when using the average were $88.35\%$ and $71.27\%$, random values from uniform distribution were $71.35\%$ and $69.28\%$, and random values from a normal distribution were $84.90\%$ and $71.84\%$ for valence and activation respectively. The multiple comparison analysis showed that both MLR and NN were significantly better at predicting the change in activation compared with any of the benchmarks. For the valence predictions the performance of the NN were significantly better than any of the benchmarks, while the MLR were only significantly better than the two random predictions.

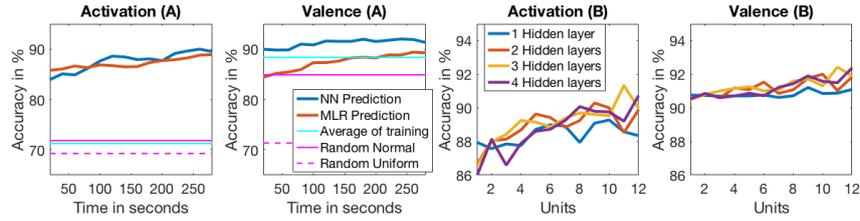

Figure 2: (A) show the accuracies as a function of the time included to make the prediction, (B) shows the accuracies as a function of the the number of units in each layer.

In Figure 2B the accuracy of the NN models' predictions (using 260 seconds windows to extract the EDR, HR, and HRV) is shown for the different number of hidden units, for each of the number of hidden layers. Table 1 shows the accuracy of the best performing NN and MLR models for the different feature-sets.

Table 1: The table shows the accuracy of the best performing NN and MLR for each feature-set, the best predictions for valence and activation are boldfaced

| feature-set | Activation NN | Activation MLR | Valence NN | Valence MLR | feature-set | Activation NN | Activation MLR | Valence NN | Valence MLR | feature-set | Activation NN | Activation MLR | Valence NN | Valence MLR |
|---|---|---|---|---|---|---|---|---|---|---|---|---|---|---|
| EDR | 90.46 | 88.86 | 91.95 | 88.95 | HR | 86.55 | 85.39 | 90.33 | 86.19 | SCL | 86.80 | 86.53 | 92.45 | 84.24 |
| HRV | 81.61 | 80.28 | 89.88 | 80.41 | EDR, SCL | 89.34 | 88.90 | 92.20 | 89.40 | EDR, HR | 90.09 | 88.67 | 92.37 | 89.23 |
| EDR HRV | 91.25 | 88.59 | 92.21 | 88.75 | SCL HR | 90.35 | 86.71 | 92.04 | 86.89 | SCL HRV | 90.04 | 86.79 | 92.10 | 84.22 |
| HR HRV | 86.80 | 85.18 | 91.33 | 86.14 | EDR SCL HR | 90.45 | 88.62 | 92.41 | 89.47 | EDR SCL HRV | 90.99 | 88.78 | 92.48 | 89.30 |
| SCL HR HRV | 90.41 | 86.44 | **92.74** | 86.75 | EDR, HR, HRV | **91.35** | 88.53 | 92.43 | 89.40 | EDR SCL HR HRV | 90.59 | 88.41 | 92.64 | 89.07 |

## 4 Discussion and conclusion

The purpose of the study was to examine if changes in affective states could be predicted, subject independently, using physiological features. The highest achieved accuracy was $92.74\%$ and $91.35\%$ when predicting changes in valence and activation, respectively. Both MLR and NN models, performed significantly better than the formulated benchmarks in the activation dimension. No significant difference between the two models were found in the activation or in the valence dimension. But the NN-model valence predictions were significantly better than all the benchmarks, while this was not the case for the MLR model.

Using longer temporal windows when computing features increased the accuracy of the predicted changes in activation for both MLR and NN, and the valence for the MLR model. This could suggests that subjects base their self-reported annotations on how they are have been feeling over a longer period of time, or that there is a delay between physiological to psychological responses to stimuli. In conclusion, it was found that predicting changes in affective states from physiological signals is possible, where only the NN outperformed the benchmarks in both dimensions.

## 5 Future work

The present study used a specific, baseline-stressor-music sequence of conditions in order to cover the entire affective space. To generalize outside this kind of laboratory conditions further experimentation is needed, to ensure the model is *universal*. In clinical environments, people tend to have abnormal physiological signals. Using a model trained on healthy patients, who are capable of communicating, in a controlled laboratory experiment, would not necessarily be the best model in a clinical settings. Hence, to confirm that continuously monitoring of affective states on patients, in different conditions, is possible, clinical studies should be performed. We have shown that the change in affective state can be predicted, however it should be noted that the position in the affective space can not be determined based on these prediction. This means that it is only possible to tell if a person is more of less pleased or activated than before, but not in general if they are pleased or feels activated. Future studies should hence try to include predictions of changes relative to place in the affective space.

**Acknowledgments**

Authors were supported in part by the Innovation Fund Denmark through the CoSound project, case number 0603-00475B. This publication only reflects the authors' views.